\documentclass[preprintnumbers,floatfix,preprint,prc]{revtex4}

\usepackage{enumerate}
\usepackage{amsmath,amssymb}  % AMS math & symbols
\usepackage{bm}               % bold math
\usepackage{amscd}            % for extensible arrows (e.g., limits)
\usepackage{graphicx}         % PostScript figures
\usepackage{hhline,multirow}  % for nicer tables
\usepackage{dcolumn}          % Align table columns on decimal point

\newcommand{\nbar}{{\overline n}}
\newcommand{\nslash}{n\hspace*{-0.22cm}\slash\hspace*{0.022cm}}
\newcommand{\nbslash}{\nbar\hspace*{-0.22cm}\slash\hspace*{0.022cm}}
\newcommand{\Tr}{{\rm Tr}}

\begin{document}
\preprint{JLAB-THY-09-1028}

\title{Hadron mass corrections in semi-inclusive	\\
	deep inelastic scattering}

\author{A.~Accardi$^{a,b}$, T.~Hobbs$^b$, W.~Melnitchouk$^b$}
\affiliation{
$^a$\mbox{Hampton University, Hampton, Virginia 23668, USA} \\
$^b$\mbox{Jefferson Lab, Newport News, Virginia 23606, USA} \\
\\
}

\begin{abstract}
We derive mass corrections for semi-inclusive deep inelastic
scattering of leptons from nucleons using a collinear factorization
framework which incorporates the initial state mass of the target
nucleon and the final state mass of the produced hadron.
The formalism is constructed specifically to ensure that physical
kinematic thresholds for the semi-inclusive process are explicitly
respected.
A systematic study of the kinematic dependencies of the mass corrections
to semi-inclusive cross sections reveals that these are even larger than
for inclusive structure functions, especially at very small and very
large hadron momentum fractions.
The hadron mass corrections compete with the experimental uncertainties
at kinematics typical of current facilities, and will be important to
efforts at extracting parton distributions or fragmentation functions
from semi-inclusive processes at intermediate energies.
\end{abstract}

\maketitle

%%%%%%%%%%%%%%%%%%%%%%%%%%%%%%%%%%%%%%%%%%%%%%%%%%%%%%%%%%%%%%%%%%%%%%%
\section{Introduction}

In recent years semi-inclusive deep inelastic scattering (SIDIS)
has received much attention as a tool to investigate various aspects
of hadron structure, such as the flavor dependence of the nucleon's
parton distribution functions, both unpolarized and polarized,
through flavor tagging of hadrons in the final state.
Observation of the momentum distribution of produced hadrons also
allows access to the largely unexplored transverse momentum dependent
parton distributions, which reveal a much richer landscape of the spin
and momentum distribution of quarks in the nucleon, and which are the
subject of increasingly greater focus at modern facilities such as
Jefferson Lab.

At high energies the scattering and hadronization components of the
SIDIS process factorize and the cross section can be represented
as a product of parton distribution and fragmentation functions.
In practice, however, experiments are often carried out at few-GeV
energies with $Q^2$ as low as 1~GeV$^2$, suggesting that $1/Q^2$ power
corrections must be controlled in order to determine the applicability
of partonic analyses of the data.

One of the standard finite-$Q^2$ corrections that must be applied in
analyses of {\em inclusive} deep inelastic scattering (DIS) data is
target mass corrections (TMCs) \cite{Schienbein}.
Kinematical in origin, TMCs arise from leading twist operators in QCD,
but enter as $1/Q^2$ corrections to structure functions \cite{Nachtmann}.
They are especially egregious at high values of the Bjorken scaling
variable $x_B$, even at relatively large $Q^2$, and are crucial for
reliable extractions of parton distributions in this region.
To date, however, the phenomenology of TMCs has not been systematically
considered in SIDIS, and we do so in this paper.

Target mass corrections in inclusive DIS have usually been formulated
within the operator product expansion, in which the subleading $1/Q^2$
corrections arise from twist-two operators involving derivative
insertions into quark bilinears
\cite{GP,Tung,Matsuda,Piccione,Blumlein,KretzerOPE,Steffens}.
Unfortunately, this method cannot be rigorously extended to the
production of hadrons in the final state.
An alternative approach to computing TMCs makes use of the collinear
factorization (CF) framework \cite{EFP,Collins,Collins2}, which has
recently been used in both unpolarized \cite{KretzerCF,AOT,AQ} and
polarized \cite{AM} inclusive DIS.
Because here one works directly in momentum space, the method can be
readily extended to SIDIS.
In contrast to inclusive DIS, where the only mass scale entering the
problem is that of the target hadron, in SIDIS finite-$Q^2$ corrections
arise from both the target mass and the mass of the produced hadron.
For generality we shall refer to their combined effects as
``hadron mass corrections'' (HMCs).

Hadron mass corrections in SIDIS at finite-$Q^2$ kinematics in
CF were considered previously in Refs.~\cite{Albino,Mulders} in
different collinear frames.
Albino {\em et al.} \cite{Albino} studied the effects of the final
state hadron mass, but did not consider the effects of the target mass.
Mulders \cite{Mulders} derived corrections due to both target and
produced hadron mass, but did not discuss the phenomenological
consequences.  Neither of these, however, addressed problems related
to kinematic thresholds.

In this work we use the CF framework to derive the mass corrections to
the SIDIS cross section at finite $Q^2$, and systematically investigate
their implications at kinematics relevant to current experiments.
The formalism is constructed specifically to ensure that physical
kinematic thresholds for the semi-inclusive process are explicitly
respected.
In Sec.~\ref{sec:semi-intro} we review the collinear formalism and
discuss its application to semi-inclusive hadron production.
To expose the origin of the corrections we work at leading order
in $\alpha_s$; next-to-leading order effects can be included in
subsequent analyses.
In Sec.~\ref{sec:results} we explore the relative importance of the
HMCs numerically, and evaluate the size of the corrections in the
cross sections and fragmentation functions at various kinematics.
To assess their possible impact on data analyses, we also compare
the magnitude of the HMCs at kinematics typical of modern facilities,
such as Jefferson Lab and HERMES, with experimental errors from
recent experiments.
Finally, in Sec.~\ref{sec:conclusion} we summarize our results
and outline avenues for future developments of this work.
A discussion of the formulation of HMCs in different collinear
frames is presented in Appendix~\ref{sec:app}.

%%%%%%%%%%%%%%%%%%%%%%%%%%%%%%%%%%%%%%%%%%%%%%%%%%%%%%%%%%%%%%%%%%%%%%%
\section{Semi-inclusive scattering at finite $Q^2$}
\label{sec:semi-intro}

We begin the discussion of SIDIS at finite values of the photon
virtuality $Q^2$ by defining the relevant kinematics and momentum
variables in a collinear frame, and introduce the hadronic tensor
computed in a covariant parton model.
Collinear factorization is then performed in the leading order
approximation in which the produced hadron is effectively collinear
with the scattered parton, which more directly reveals the effects
of hadron masses on the cross sections and fragmentation functions.

% .......................................................................
\subsection{External kinematics} 

The four-momenta of the target nucleon ($p$), virtual photon ($q$)
and produced hadron $h$ ($p_h$) can be decomposed in terms of
light-cone unit vectors $n$ and $\nbar$ as \cite{EFP}
\begin{subequations}
\begin{eqnarray}
p^\mu   &=& p^+\, \nbar^\mu 
         + \frac{M^2}{2 p^+}\, n^\mu\ ,		\\
q^\mu   &=& - \xi p^+\, \nbar^\mu 
         + \frac{Q^2}{2\xi p^+}\, n^\mu\ ,	\label{eq:kin1b} \\
p_h^\mu &=& \frac{\xi m_{h\perp}^2}{\zeta_h Q^2} p^+\, \nbar^\mu
	 + \frac{\zeta_h Q^2}{2 \xi p^+}\, n^\mu 
	 + p_{h\perp}^{\,\mu} \ ,		\label{eq:kin1c}
\end{eqnarray}
\label{eq:kinematics}%
\end{subequations}%
where $M$ is the target nucleon mass, $Q^2=-q^2$, and the light-cone
vectors satisfy $n^2 = \nbar^2 = 0$ and $n \cdot \nbar=1$.
Here we define light-cone components of any four-vector $v$ by
$v^+ = v \cdot \nbar = (v_0 + v_z)/\sqrt{2}$ and
$v^- = v \cdot n     = (v_0 - v_z)/\sqrt{2}$. 
The momenta $p$ and $q$ are chosen to lie in the same plane as $n$
and $\nbar$, as for inclusive DIS. We call this the $(p,q)$ collinear
frame; other possible choices are discussed and compared in Appendix~A.
The nucleon plus-momentum $p^+$ can be interpreted as a parameter for
boosts along the $z$-axis, connecting the target rest frame to the
infinite-momentum frame; the target rest frame ($p^+=M/\sqrt{2}$) and
the Breit frame ($p^+=Q/(\sqrt{2}\xi$)) are part of this family of frames. 
The transverse momentum four-vector of the produced hadron
$p_{h\perp}^{\,\mu}$ satisfies
$p_{h\perp} \cdot n = p_{h\perp} \cdot \nbar = 0$,
and we define the transverse mass squared as
$m_{h\perp}^2 = m_h^2 - p_{h\perp}^2$,
where $m_h$ is the produced hadron mass, and the transverse
four-vector squared is
$p_{h\perp}^2 = -\bm{p}_{h\perp}^{\,2}$.

In the chosen collinear frame the variable $\xi = - q^+ / p^+$ 
defined in Eq.~\eqref{eq:kin1b} coincides with the finite-$Q^2$
Nachtmann scaling variable \cite{Nachtmann,Greenberg},
\begin{align}
\xi = \frac{2 x_B}{1 + \sqrt{1 + 4 x_B^2 M^2/Q^2}}\ ,
\label{eq:xi}
\end{align}
which in the Bjorken limit ($Q^2 \to \infty$ at fixed $x_B$)
reduces to the Bjorken scaling variable $x_B=Q^2/2p\cdot q$.
The scaling fragmentation variable $\zeta_h = p_h^- / q^-$ defined
in Eq.~\eqref{eq:kin1c} is related to the fragmentation invariant
$z_h = p_h \cdot p / q \cdot p$ by
\begin{subequations}
\label{eq:zeta_h}
\begin{align}
\zeta_h
  & = \frac{z_h}{2} \frac{\xi}{x_B}
  \left( 1 + \sqrt{1 - \frac{4 x_B^2 M^2 m_{h\perp}^2}{z_h^2\ Q^4}} 
  \right)\, ,
\label{eq:zeta_h1}
\end{align}
and the positivity of the argument in the radical in
Eq.~(\ref{eq:zeta_h1}) is ensured by the condition $E_h \geq m_{h\perp}$,
which imposes
\begin{align}
z_h \geq z_h^{\rm min} = 2 x_B \frac{M m_h}{Q^2} \ .
\end{align}
One can also define $\zeta_h$ in terms of the invariant
$\eta_h = 2p_h \cdot q / q^2$ by
\begin{align}
\zeta_h
  & = \frac{\eta_h}{2} 
  \left( 1 + \sqrt{1 + \frac{4 m_{h\perp}^2}{\eta_h^2\ Q^2}} 
  \right)\ ,
\label{eq:eta_h2}
\end{align}
\end{subequations}
which is convenient for discriminating between the target and current
fragmentation hemispheres in hadron production.
Note that in the target rest frame $z_h = E_h/\nu$ is the usual
ratio of the produced hadron to virtual photon energies. 
In the Breit frame $\eta_h=p_{hz}/q_z$ is the ratio of the longitudinal 
components of the hadron and photon energies, which can be used to
define the current ($\eta_h>0$) and target ($\eta_h<0$) hemispheres
for hadron production.
In the Bjorken limit one has $\zeta_h \to z_h \to \eta_h$.

Conservation of four-momentum and baryon number impose an upper limit
on the $x_B$ variable,
\begin{gather}
  x_B \leq \left( 1 + {m_h^2 + 2 M m_h \over Q^2} \right)^{-1}
      \equiv x_B^{\rm max}\, ,
\label{eq:xBlim}
\end{gather}
which corresponds to the exclusive production of a nucleon and
a hadron $h$ in the final state.
Similarly the limits on the fragmentation variable $\zeta_h$ are
given by
\begin{gather}
  \frac{\xi}{1-\xi} \frac{M^2}{Q^2} \leq \zeta_h 
    \leq 1 + \xi \frac{M^2}{Q^2}\, ,
\label{eq:zetahlim}
\end{gather}
where the lower limit corresponds to diffractive production of the
hadron $h$, and the upper limit reflects the fragmentation threshold,
which approaches unity in the Bjorken limit.

% .......................................................................
\subsection{Parton kinematics in collinear factorization}

\begin{figure}[t]
\includegraphics[height=8cm]{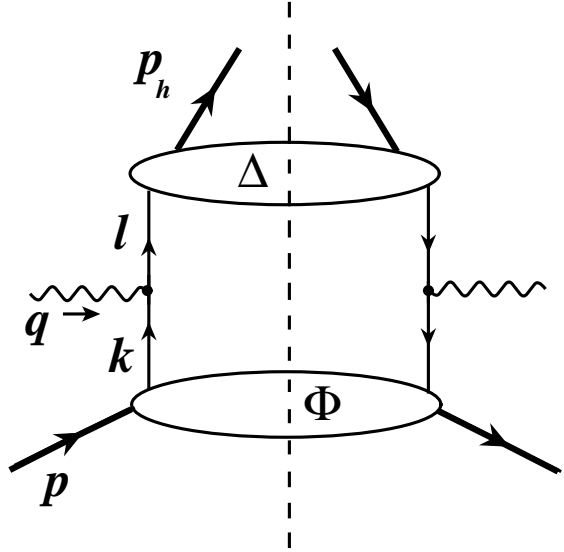}
\caption{Kinematics of semi-inclusive deep inelastic lepton--nucleon
	scattering at leading order, producing a final state hadron $h$.
	The momenta of the target nucleon ($p$), virtual photon ($q$),
	incident ($k$) and scattered quarks ($l$), and the produced
	hadron ($p_h$) are labeled explicitly, with $\Phi$ and $\Delta$
	denoting the correlators relevant to the quark distribution
	and fragmentation functions.  The vertical dashed line
	represents the cut of the forward amplitude.}
\label{fig:1}
\end{figure}

At the partonic level the SIDIS process at leading order in the strong
coupling constant $\alpha_s$ is illustrated in Fig.~\ref{fig:1}.
It proceeds through the scattering from a quark carrying a light-cone
momentum fraction $x = k^+ / p^+$, which then fragments to a hadron $h$
carrying a light-cone momentum fraction $z = p_h^-/l^-$, where $k$ and
$l$ are the four-momenta of the initial and scattered quarks.
At higher orders the hard scattering can also take place from a gluon,
and additional partons can be created in the collision.

The parton momenta $k$ and $l$ can be parametrized in terms of the
light-cone vectors $n$ and $\nbar$ as
\begin{subequations}
\begin{eqnarray}
k^\mu &=& xp^+\, \bar{n}^\mu
       + \frac{k^2 + k_\perp^2}{2 x p^+}\, n^\mu
       + k_\perp^\mu\ ,				\\
l^\mu &=&  \frac{l^2 + l_\perp^2}{2 p_h^-/z}\, \bar{n}^\mu
       + \frac{p_h^-}{z}\, n^\mu
       + l_\perp^\mu\ ,
\end{eqnarray}
\end{subequations}
with the parton transverse momentum four-vectors $k_\perp$ and
$l_\perp$ orthogonal to $n$ and $\nbar$. 
In collinear factorization the hard scattering amplitude is expanded
around on-shell and collinear momenta $\tilde k$ and $\tilde l$,
\begin{subequations}
\begin{eqnarray}
  \tilde k^\mu &=& xp^+\, \bar{n}^\mu
       + \frac{\tilde k^2}{2 x p^+}\, n^\mu \, \\
  \tilde l^\mu &=&  \frac{\tilde l^2 + p_{h\perp}^2/z^2}{2 p_h^-/z}\,
		    \bar{n}^\mu
       + {p_h^- \over z}\, n^\mu
       + {p_{h\perp}^\mu \over z}\ ,
\end{eqnarray}
\end{subequations}
where the initial and final collinear parton ``masses'' $\tilde k^2$
and $\tilde l^2$ are kept for generality.

Defining the invariant $\hat x = -q^2/2\tilde k\cdot q$ as the partonic
analog of the Bjorken variable $x_B$, at finite $Q^2$ one has
\begin{align}
\hat x = {\xi \over x}
         \left( 1 + {x \over \xi} {\tilde k^2 \over Q^2} \right)\, .
\label{eq:xf}
\end{align}
Using the methods described in Ref.~\cite{AQ} one can show that for
SIDIS cross sections integrated over $p_{h \perp}$, $\hat x$ is
constrained to be in the range
\begin{align}
  1 + \frac{m_h^2}{\zeta_h Q^2}
    - \frac{\tilde k^2}{Q^2}
      \left( 1 - \frac{\xi m_h^2}{x \zeta_h Q^2} \right)
  \leq \, \frac{1}{\,\hat x\,} \, \leq 
  \frac{1}{x_B}
    \left( 1 - x_B \frac{2 M m_h + \tilde k^2}{Q^2} \right)\ ,
\label{eq:xflimits}
\end{align}
where the lower limit arises from the minimum of the current jet mass,
and the upper limit corresponds to collinear spectators with minimal mass.
These limits agree with the limit on $x_B$ in Eq.~\eqref{eq:xBlim} for
any $\tilde k^2 \geq x (\zeta_h-1)Q^2/\xi$.
For the fragmentation process one finds analogous limits on $\zeta_h$,
\begin{align}
  \zeta_h\ \leq\ {1 \over z}\, \zeta_h\
	   \leq\ 1 + {\xi \over x} {\tilde k^2 \over Q^2} \ ,
\label{eq:wlimits}
\end{align}
which agrees with the limit in Eq.~\eqref{eq:zetahlim},
provided that $\tilde k^2 \leq x M^2$. 
The requirement that the collinear parton masses be independent
of the parton momentum ({\it viz.}, independent of $x$) implies
$\tilde k^2 \leq 0$.
Combined with the above lower limit on $\tilde k^2$, this naturally
leads to a collinear expansion around a massless initial state parton,
$\tilde k^2 = 0$.

The choice of $\tilde l^2$ is made by considering the cross section
at leading order in $\alpha_s$.
Four-momentum conservation for the hard scattering, together with
the choice $\tilde k^2 = 0$, leads to the relations
$x = \xi (1 + \tilde l^2/Q^2) \equiv \xi_h$ and $z = \zeta_h$.
Clearly $z$ falls within the kinematic limits \eqref{eq:wlimits}.
However, in order for $x$ to respect the limits \eqref{eq:xflimits}
we choose $\tilde l^2 = m_h^2/\zeta_h$, in which case
\begin{align} 
\xi_h\ =\ \xi\, \left( 1 + {m_h^2 \over \zeta_h\, Q^2} \right)\, .
\label{eq:xi_h}
\end{align}
While larger values of $\tilde l^2$ would also allow $x$ to fall
within the limits \eqref{eq:xflimits}, this choice is the closest to
the physical quark mass.

We stress that our prescription for the collinear parton masses
$\tilde k^2$ and $\tilde l^2$ is dictated by the external kinematic
limits in Eqs.~\eqref{eq:xBlim} and \eqref{eq:zetahlim}, which are
independent of the parton model and collinear factorization
approximations.
As discussed in Refs.~\cite{AQ,CRS}, this is crucial when considering
cross sections close to the kinematic limits, such as at large $x_B$
or large $z_h$.
However, as we shall see in the next section, the SIDIS cross section
can also receive non-negligible corrections at small $x_B$ since
$\xi_h > \xi \approx x_B$.
This is qualitatively different from the behavior of the target mass
corrections in inclusive DIS, which are always suppressed at small $x_B$
\cite{AQ}.

% .......................................................................
\subsection{Hadron tensor and cross section at leading order}

In collinear factorization the hadron tensor at leading order, to
which we restrict the rest of our analysis, can be written as
\begin{align}
2 M W^{\mu\nu}(p,q,p_h)
= \sum_q e_q^2 \int d^4k\ d^4l\ \delta^{(4)}(\tilde k + q - \tilde l)\
  {\rm Tr}[ \Phi_q(p,k)\, \gamma^{\mu}\,
	    \Delta_q^h(l,p_h)\, \gamma^{\nu} ]\ ,
\label{eq:Wmunu}
\end{align}
where the sum is taken over quark flavors $q$, and the correlators
$\Phi_q$ and $\Delta_q^h$ encode the relevant quark distribution and
fragmentation functions, respectively \cite{Collins,Collins2,Mulders}.
According to our prescription for the collinear momenta, the
$\delta$-function depends on the collinear momenta $\tilde k$
and $\tilde l$, so that integrations over $dk^-\, d^2k_{\perp}$
and $dl^+\, d^2l_{\perp}$ act directly on the correlators $\Phi$
and $\Delta$.
The leading twist part of the cross section can then be extracted 
by retaining the $\nbslash$ and $\nslash$ components in the Dirac
structure expansion of the integrated correlators,
\begin{subequations}
\begin{align}
  \int dk^- d^2k_{\perp}\, \Phi_q(p,k)
	& = \frac12 f_q(x) \nbslash + \ldots\ ,		\\
  \int dl^+ d^2l_{\perp}\, \Delta_q^h(l,p_h)
	& = \frac12 D_q^h(z) \nslash + \ldots\ ,
\end{align}
\end{subequations}
where the dots indicate contributions of higher twist \cite{TMD}.
The nonperturbative quark distribution function $f_q(x)$ and
quark-to-hadron fragmentation function $D_q^h(z)$ are explicitly
defined as
\begin{subequations}
\begin{align}
f_q(x)
    &= \frac12 \int dk^- d^2k_\perp\, \Tr\left[ \gamma^+ 
    \Phi_q(p,k)\right]_{k^+=xp^+}\ 		\nonumber \\
    & \stackrel{\text{LC}}{=}
    \frac{1}{2} \int \frac{dw^-}{2\pi} e^{i xp^+ w^-}
    \langle N | \overline{\psi}_q(0)\, \gamma^+\, \psi_q(w^- n)
    | N \rangle\ , \\
D_q^h(z)
    &= \frac{z}{2} \int dl^+ d^2l_\perp\,
	   \Tr\left[\gamma^-\Delta_q^h(l,p_h)\right]_{l^-=p_h^-/z}
						\nonumber \\
    &\stackrel{\text{LC}}{=}
    \frac{z}{2} \sum_X \int \frac{dw^+}{2\pi} e^{i(p_h^-/z)w^+}
    \langle 0 | \psi_q(w^+ n) |h, X \rangle
    \langle h,X | \overline{\psi}_q(0) \gamma^-| 0 \rangle\ , 
\end{align}
\end{subequations}
where ``LC'' denotes use of the light-cone gauge, and the
fragmentation function is normalized such that
$\sum_h \int_0^1 dz\, z\, D_q^h(z) = 1$ \cite{Mulders}.

% Here one implicitly assumes that $k^-, k_\perp \ll \Lambda$ and
% $l^+, l_\perp \ll \Lambda$, where $\Lambda$ is a hard momentum scale,
% which in SIDIS can be taken to be $Q$.

From Eq.~(\ref{eq:Wmunu}) the energy-momentum conserving
$\delta$-function can be decomposed along the plus, minus,
and transverse components of the light-cone momentum.
The plus and minus components yield a product of $\delta$-functions
that fix $x = \xi_h$ and $z = \zeta_h$, while the transverse component
constrains the transverse momentum of the scattered quark to vanish,
which restricts the produced hadrons to be purely longitudinal,
$p_{h\perp} = z\, l_\perp=0$.
Hadrons with nonzero transverse momentum can be generated from
higher order perturbative QCD processes, or from intrinsic
transverse momentum in the parton distribution functions, as in
the case of transverse momentum dependent distributions \cite{TMD},
but are not considered in this work.
The resulting hadron tensor in the presence of hadron mass effects,
\begin{align}
2M W^{\mu\nu}(p,q,p_h)
= \frac{\zeta_h}{4} \sum_q e_q^2\ 
  \delta^{(2)}(\bm{p}_\perp)
  \Tr \left[ \nbslash\gamma^{\mu}\nslash\gamma^{\nu} \right]
  f_q(\xi_h) D_q^h(\zeta_h)\, ,
\label{eq:TMC_hadtens}
\end{align}
is then factorized into a product of parton distribution and
fragmentation functions evaluated at the finite-$Q^2$ scaling
variables $\xi_h$ and $\zeta_h$, instead of $x_B$ and $z_h$ as
would be obtained in the massless case, and recovered from
Eq.~\eqref{eq:TMC_hadtens} in the Bjorken limit.
Note that this prescription is the same as that used in Ref.~\cite{AQ}
when discussing inclusive DIS in the presence of jet mass corrections,
and is close in spirit to that advocated in Ref.~\cite{CRS}, where the
trace is calculated as in the massless case, but overall parton
momentum conservation respects the external kinematics.

Finally, the SIDIS cross section is computed by contracting the
hadron tensor with an analogous lepton tensor \cite{TMD}, leading to
\begin{align}
\sigma\ \equiv\ \frac{d\sigma}{dx_B\, dQ^2\, dz_h}
&=\ \frac{2\pi\alpha_s^2}{Q^4} \frac{y^2}{1-\varepsilon}
    \frac{d\zeta_h}{dz_h}
    \sum_q e_q^2\, f_q(\xi_h,Q^2)\, D_q^h(\zeta_h,Q^2)\ ,
\label{eq:dsigTMC}
\end{align}
where the dependence of the functions on the scale $Q^2$ is made
explicit, and the Jacobian
$d\zeta_h/dz_h = (1 - M^2\xi^2/Q^2)
	       / (1-\xi^2 M^2 m_h^2/\zeta_h^2 Q^4)$.
In Eq.~\eqref{eq:dsigTMC} the variable $y$ defined as
$y = p\cdot q / p\cdot p_\ell$, where $p_\ell$ is the lepton momentum,
represents the fractional energy transfer from the lepton to the hadron in the
target rest frame ($y=\nu/E$, with $E$ the lepton energy), and
$\varepsilon = (1 - y - y^2 \gamma^2/4)
	     / (1 - y + y^2 [1/2+\gamma^2/4])$
is the ratio of longitudinal to transverse photon flux, with
$\gamma^2 = 4 x_B^2 M^2/Q^2$.
The cross section differential in $\eta_h$ can be obtained using
$d\zeta_h/d\eta_h = 1/(1+m_h^2/\zeta_h^2 Q^2)$ instead of
$d\zeta_h/dz_h$.
It is interesting to observe that since $\xi_h$ depends explicitly
on $m_h$ and $\zeta_h$ depends on $z_h$ and $x_B$, at finite $Q^2$
the scattering and fragmentation parts of the cross section
(\ref{eq:dsigTMC}) are not independent.

As a final remark we note that at the maximum allowed $x_B$ for
SIDIS, Eq.~(\ref{eq:xBlim}), the value of $\xi_h$ is smaller than  
$\xi_h (x_B=x_B^{\rm max}) < 1$.
As in the case of inclusive DIS \cite{AQ}, the SIDIS cross section
therefore does not vanish as $x_B \to x_B^{\rm max}$, which is a
manifestation of the well-known threshold problem \cite{Schienbein}.
On the other hand, from Eq.~\eqref{eq:wlimits} the fragmentation
variable $\zeta_h \leq 1$, and no threshold problem appears in the 
fragmentation function since $D(\zeta_h) \to 0$ as $\zeta_h \to 1$.

In the next section we shall examine the phenomenological consequences
of the finite-$Q^2$ rescaling of the SIDIS cross section numerically.

%%%%%%%%%%%%%%%%%%%%%%%%%%%%%%%%%%%%%%%%%%%%%%%%%%%%%%%%%%%%%%%%%%%%%%%
\section{Hadron Mass Corrections}
\label{sec:results}

Using the hadron mass corrected expressions for the SIDIS cross
section derived above, % in Eq.~\eqref{eq:dsigTMC},
we next explore the dependence of the cross sections and fragmentation
functions on the fragmentation variable $z_h$, for various $x_B$ and
$Q^2$ values and for different final state hadron masses.
We then compare the relative size of the HMCs with the experimental
uncertainties from recent SIDIS experiments at Jefferson Lab and the
HERMES Collaboration, as well as with higher energy data from the
European Muon Collaboration (EMC) and HERA.

\begin{figure}[t]
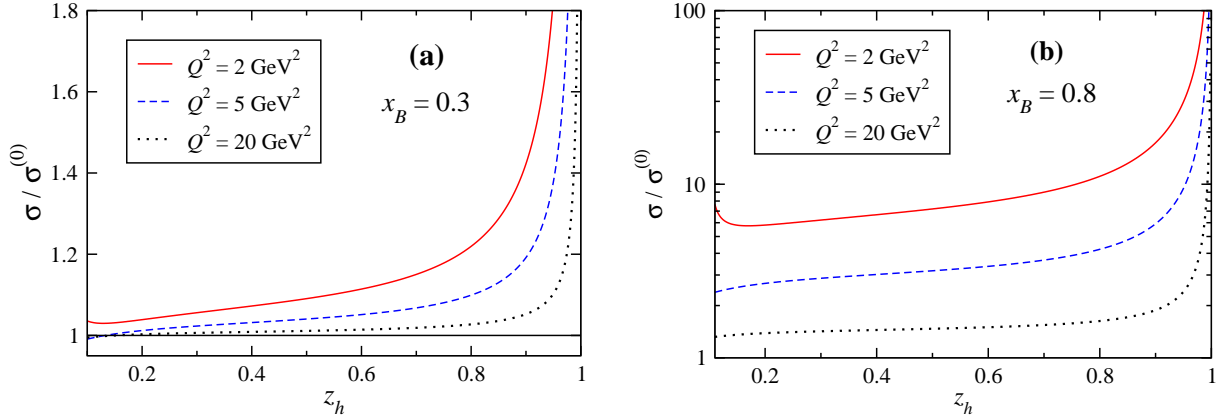

\includegraphics[height=5.5cm]{Fig2a.eps}\ \ \ \ \ 
\includegraphics[height=5.5cm]{Fig2b.eps}
\caption{Ratio of cross sections $\sigma/\sigma^{(0)}$ for
        semi-inclusive charged-pion production (($\pi^+ + \pi^-)/2$)
        as a function of $z_h$ at several $Q^2$ values for
        (a) $x_B = 0.3$ and (b) $x_B = 0.8$.}
\label{fig:2}
\end{figure}

% .......................................................................
\subsection{HMC phenomenology}

To illustrate most directly the effects of the HMCs, in
Fig.~\ref{fig:2} we consider charged pion production (average of
$\pi^+$ and $\pi^-$) and plot as a function of $z_h$, for different
$x_B$ and $Q^2$, the ratio of the full cross section $\sigma$ in
Eq.~(\ref{eq:dsigTMC}) to the cross section $\sigma^{(0)}$,
defined by taking the massless limit for the scaling variables
$\sigma^{(0)} \equiv \sigma(\xi_h \to x_B, \zeta_h \to z_h)$ and
setting $d\zeta_h/dz_h=1$.
For the numerical computations we use the leading order CTEQ6L
parton distributions \cite{PDF} and the KKP leading order
fragmentation functions \cite{KKP}, unless otherwise specified.
The ratio at $x_B = 0.3$ in Fig.~\ref{fig:2}(a) is enhanced by
$\leq 20\%$ at $Q^2=2$~GeV$^2$ for $z_h \lesssim 0.7$, but rises
dramatically as $z_h \to 1$.
The effect is naturally smaller at higher $Q^2$ values, but the
rise at high $z_h$ is a common feature for all kinematics.
The same ratios at $x_B = 0.8$ in Fig.~\ref{fig:2}(b) show
approximately an order of magnitude larger overall effect
(note the logarithmic scale!).

\begin{figure}[t]
\includegraphics[height=6cm]{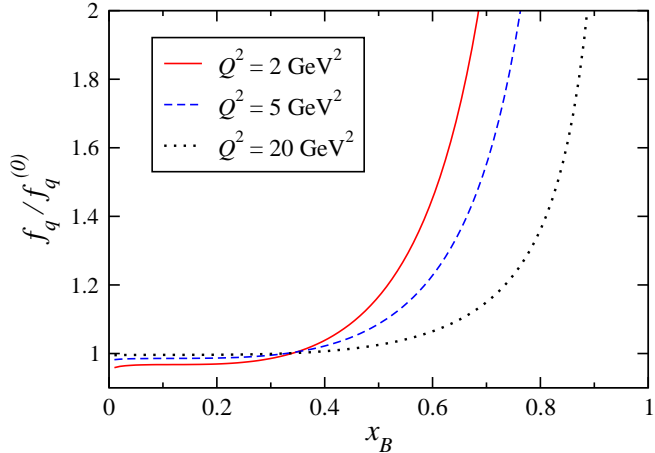}
\caption{Ratio of the hadron mass corrected isoscalar parton
	distribution function $f_q(\xi_h)$ for $q=u+d$ to the
	massless limit distribution $f_q^{(0)}$ as a function
	of $x_B$, for $m_h = m_\pi$ and $\zeta_h = 0.2$. \\ \\ \\}
\label{fig:3}
\end{figure}

The small upturn in the ratios at low $z_h$ for the lowest $Q^2$ in
Fig.~\ref{fig:2} can be understood from the interplay between the
finite-$Q^2$ kinematics and the shape of the fragmentation function.
Assuming the fragmentation function is smooth, one can expand the
ratio of corrected to uncorrected functions in a Taylor series as
\begin{equation}
\frac{D(\zeta_h)}{D(z_h)} \approx
1 + \frac{D^\prime(z_h)}{D(z_h)} (\zeta_h - z_h)\, .
%    \right|_{\zeta_h = z_h} 
\label{eq:TMCdif}
\end{equation}
The $z_h$ dependence of the HMCs arising in the fragmentation function 
is mostly determined by the negative shift in the fragmentation 
variable ($\zeta_h-z_h$) and by the local rate of change over $z_h$
of the fragmentation function.
The pion fragmentation function generally behaves as a negative power
of $z_h$ at small $z_h$, and the negative slope drives the ratio of
corrected to uncorrected fragmentation functions upward as
$z_h \to z_h^{\rm min}$, where $|\zeta_h-z_h|$ is maximum.
For kaons and protons the slope of the form factor can be positive,
which would suppress the mass corrected cross section in the vicinity
of $z_h^{\rm min}$. 
In the limit $z_h \to 1$, on the other hand, the ratio
$\sigma/\sigma^{(0)}$ becomes divergent for any kinematics and any
hadron species because the cross section $\sigma^{(0)} \propto D(z_h)$
vanishes, while the rescaled cross section remains finite.

At very small values of $z_h$ the factor $(1+m_h^2/\zeta_h Q^2)$ in
the definition of $\xi_h$ in Eq.~(\ref{eq:xi_h}) can render $\xi_h$
larger than $x_B$, suppressing the $\xi_h$-rescaled parton distributions
relative to their asymptotic limit and driving $\sigma/\sigma^{(0)}$
slightly below unity.
As discussed below, for heavier hadrons this effect will be more
pronounced. 
The effect of the $\xi_h$ rescaling on the SIDIS cross section is 
illustrated explicitly in Fig.~\ref{fig:3}, where we show the ratio 
of the isoscalar parton distribution functions $f_q$, $q=u+d$,
with [$f_q=f_q(\xi_h)$] and without [$f_q^{(0)}=f_q(x_B)$] hadron mass
corrections, as a function of $x_B$ for $\zeta_h=0.2$ and $m_h=m_\pi$.
At $Q^2 = 2$~GeV$^2$ the mass corrected parton distribution is
several times larger than the uncorrected one at $x_B=0.8$, and
even at $Q^2 = 20$~GeV$^2$ the HMC is some 50\%, with the effect
increasing dramatically as $x_B \to 1$.
This sharp rise is analogous to that in inclusive DIS \cite{AQ}, and
arises from $\xi_h$ being smaller than $x_B$ when the latter is large.
This is responsible for the large overall magnitude of the corrections
in Fig.~\ref{fig:2}(b) compared with those at $x_B=0.3$.
In contrast, the $\xi_h$ rescaling effect becomes quite small at
$x_B \lesssim 0.3$ for all the $Q^2$ considered, and in fact drives
the ratio below unity, as discussed above.

\begin{figure}[t]
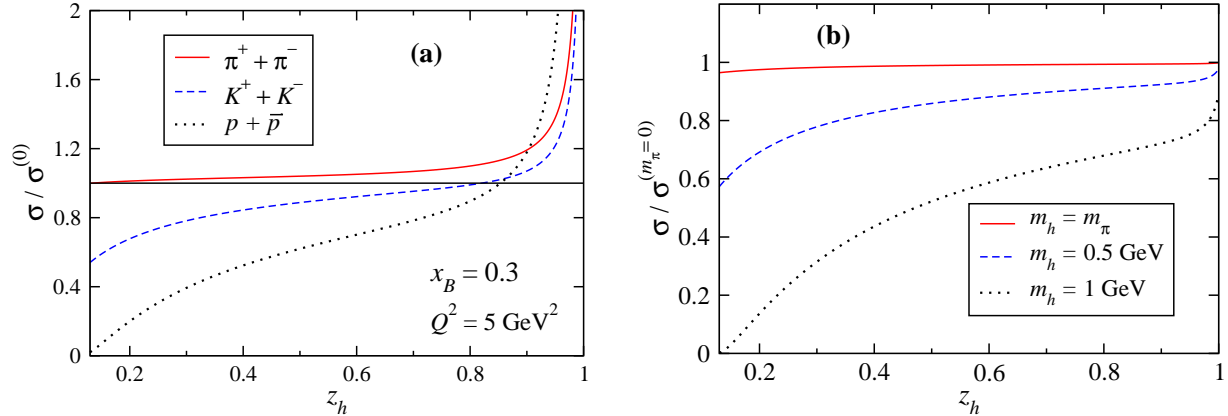

\includegraphics[height=5.5cm]{Fig4a.eps}\ \ \ \ \
\includegraphics[height=5.5cm]{Fig4b.eps}
\caption{(a) Dependence of the ratio of SIDIS cross sections
	$\sigma/\sigma^{(0)}$ with and without HMCs for different
	produced hadrons, $h=\pi^+ + \pi^-$, $K^+ + K^-$ or
	$p + \bar p$.
	(b) Ratio of cross sections for $h=\pi^+ + \pi^-$ for
	different values of the pion mass, relative to the massless
	cross section.
	In both cases the kinematics chosen are $x_B = 0.3$ and
	$Q^2 = 5~$GeV$^2$.}
\label{fig:4}
\end{figure}

The relative importance of HMCs for different produced hadron
species is illustrated in Fig.~\ref{fig:4}(a), where the
ratio $\sigma/\sigma^{(0)}$ is shown as a function of $z_h$
for $x_B = 0.3$ and $Q^2 = 5~$GeV$^2$.
Over the range $0.3 \lesssim z_h \lesssim 0.8$ the HMCs yield
an upward correction of $\lesssim 10\%$ for the pions,
but a downward correction of $\lesssim 20\%$ and $\lesssim 40\%$
for kaons and protons/antiprotons, respectively.
At lower $z_h$ the cross section ratio for the heavier hadrons
decreases dramatically because of the large suppression of
the parton distribution from the $(1+m_h^2/\zeta_h Q^2)$ factor
in $\xi_h$, which overwhelms any other small-$z_h$ effect.

Note that in Fig.~\ref{fig:4}(a) the appropriate fragmentation function
for each produced hadron species has been used, which introduces a
flavor dependence in the HMC because of the different fragmentation
function shapes for each hadron.
To isolate the effects of the hadron mass alone, in Fig.~\ref{fig:4}(b)
the ratios of cross sections computed with charged pion fragmentation
functions and masses $m_h = m_\pi$ (= 0.139~GeV), 0.5~GeV and 1~GeV
are shown relative to the cross section with $m_\pi=0$, for which
$\zeta_h = z_h \xi / x_B$.
One can see that in general increasing the hadron mass suppresses the
cross section because of the $\xi_h$ scaling, and the inversion of the
HMC hierarchy in Fig.~\ref{fig:4}(a) going from low to high $z_h$ is
due to the increasingly negative large-$z$ slope of the fragmentation
functions for kaons and protons.
While the differences at the physical pion mass are very small,
for larger hadron masses $\sim 1$~GeV the effects can be quite
significant at $z_h \lesssim 0.4$ even for $Q^2$ values of
several~GeV$^2$.

% .......................................................................
\subsection{Implications for experiments}

\begin{figure}[t]
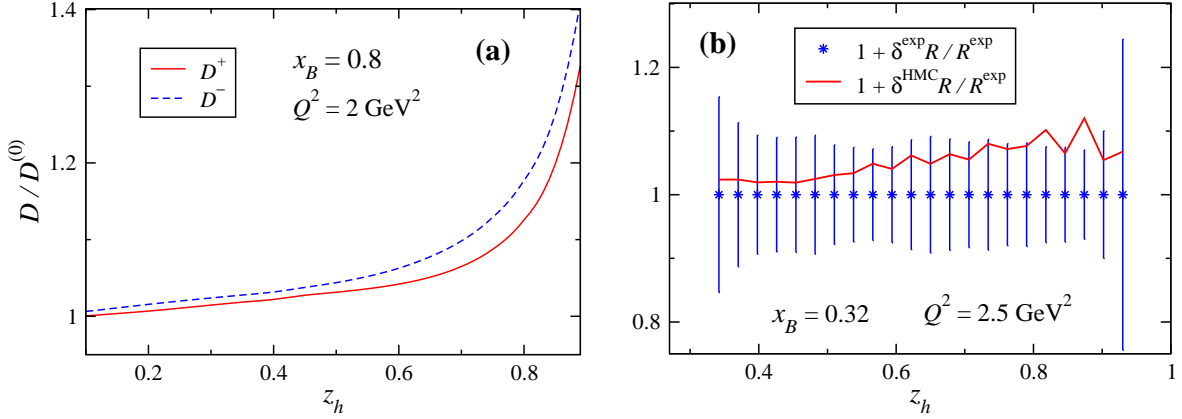

\includegraphics[height=5.5cm]{Fig5a.eps}\ \ \ \ \ 
\includegraphics[height=5.5cm]{Fig5b.eps}
\caption{(a) Ratio of hadron mass corrected to uncorrected
	fragmentation functions $D/D^{(0)}$ for favored (solid)
	and unfavored (dotted) production of $\pi^+$, for
	$x_B = 0.8$ and $Q^2 = 2$~GeV$^2$.
	(b) Comparison of the hadron mass correction
	$\delta^{\rm HMC} R$ to the ratio of unfavored to favored
	fragmentation functions $R = D^-/D^+$ with experimental errors
	$\delta^{\rm exp} R$ on $R$ from the recent Jefferson Lab
	experiment E00-108 \cite{HallC}, normalized to the central
	values of the data points.}
\label{fig:5}
\end{figure}

One of the unique capabilities of SIDIS is the ability to tag individual
quark flavors by selecting specific hadrons in the final state.
For example, because of its valence quark content, production of
$\pi^+$ is mostly sensitive to the $u$ quark, requiring only a single
$q\bar q$ pair creation from the vacuum, while $\pi^-$ reflects mostly
the $d$ quark content of the target nucleon.
This simple picture of primary fragmentation provides a good
approximation to the production mechanism only at large $z_h$,
however, and at low $z_h$ secondary fragmentation involving two or
more $q\bar q$ pair productions dilutes the direct flavor tagging.
The primary fragmentation process is parametrized by the ``favored''
fragmentation function $D^+$, describing
$u \to \pi^+$ or $d \to \pi^-$ hadronization, while the secondary
fragmentation is parametrized by the ``unfavored'' fragmentation
function $D^-$, describing $u \to \pi^-$ or $d \to \pi^+$ hadronization.

Because the $D^+$ and $D^-$ functions have rather different $z_h$
dependence, with unfavored fragmentation strongly suppressed at
large $z_h$, they will be affected differently by the hadron mass
corrections: one would expect larger HMCs for the unfavored process
since the magnitude of the slope $|dD(z_h)/dz_h|$ in
Eq.~(\ref{eq:TMCdif}) is larger for $D^-$ than for $D^+$.
In Fig.~\ref{fig:5}(a) one observes precisely this; here, we provide
an upper limit (given the choice of $x_B=0.8$ and $Q^2=2$~GeV$^2$)
to the relative size of the mass effect in $D/D^{(0)}$, which is 
universally larger in the unfavored fragmentation function.
In the numerical computations we have used the favored and 
unfavored fragmentation functions from Ref.~\cite{Sassot}.
At lower $x_B$ the correction will be smaller, although the 
qualitative features of the effect will remain.

The relevance of the HMCs to experimental data on the ratio
$R = D^-/D^+$ is expressed in Fig.~\ref{fig:5}(b), which directly 
compares the difference
$\delta^{\rm HMC} R = (D^-/D^+) - (D^-/D^+)^{(0)}$ to the
statistical uncertainty $\delta^{\rm exp} R$ in the extracted values
of $R$ from the recent Jefferson Lab experiment E00-108 \cite{HallC}
at $x_B=0.32$ and $Q^2 \approx 2.5$~GeV$^2$, with both quantities
normalized to the central values of the measured $R$ ratio.
While at small $z_h$ the HMC is relatively small compared with the
experimental errors, at large $z_h$ ($\gtrsim 0.6$) it begins to
compete with the experimental uncertainty, suggesting that the
hadron mass here poses a non-negligible effect.

\begin{figure}[t]
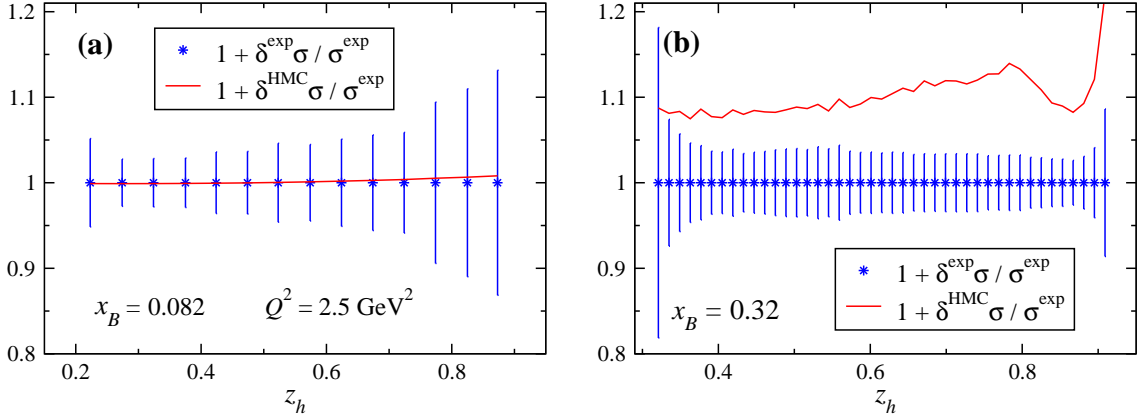

\includegraphics[height=5.5cm]{Fig6a.eps}\ \ \ \ \ 
\includegraphics[height=5.5cm]{Fig6b.eps}
\caption{Comparison of the hadron mass correction to the SIDIS cross
	section for charged hadron production, $\delta^{\rm HMC}\sigma$,
	relative to the experimental cross section,
	$\delta^{\rm exp} \sigma$, with the relative experimental
	uncertainty as a function of $z_h$ for
	(a) HERMES experiment \cite{HERMES} at $Q^2=2.5$~GeV$^2$
	and $x_B = 0.082$, and
	(b) Jefferson Lab experiment E00-108 \cite{HallC} at
	a similar $Q^2$ but at $x_B=0.32$.}
\label{fig:6}
\end{figure}

The importance of the hadron mass corrections for experimental
cross sections is examined in Fig.~\ref{fig:6}, where we compare
the calculated difference
$\delta^{\rm HMC}\sigma \equiv \sigma - \sigma^{(0)}$ with the 
experimental uncertainties $\delta^{\rm exp}\sigma$, normalized to
the central values of the cross section for charged hadron production 
from HERMES \cite{HERMES} and Jefferson Lab \cite{HallC}.
Since both experiments are dominated by the semi-inclusive production
of pions, so that $\xi_h\approx\xi$, HMCs generally produce upward 
shifts relative to data. 
The mass corrections at the HERMES kinematics in Fig.~\ref{fig:6}(a),
where $Q^2 \approx 2.5$~GeV$^2$ and $x_B = 0.082$, are generally
very small compared with the size of the experimental uncertainties.
At higher energies, HMCs to fixed-angle measurements by the EMC
\cite{EMC} at large $x_B$ values are also found to be negligible
due to suppression by $Q^2$, which increases with $x_B$.
Were these experiments conducted at smaller angles, however, it is 
likely that HMCs would become important.

On the other hand, for the Jefferson Lab experiment E00-108 \cite{HallC} 
in Fig.~\ref{fig:6}(b) at a similar $Q^2$ but larger $x_B = 0.32$, the 
mass effects are approximately $2$ times larger than the experimental
statistical  errors.
This illustrates the potentially significant impact that HMCs can have
on leading-twist analyses of SIDIS data at moderate and large $x_B$ and
low $Q^2$.
To avoid these effects one would either need to go to smaller $x_B$ or 
larger $Q^2$ values, for example afforded by the 12~GeV energy upgrade 
at Jefferson Lab.
Alternatively, since the HMCs are calculated and model independent, 
lower $Q^2$ and higher $x_B$ data will still yield useful leading twist 
information provided the mass corrections are accounted for.

Measurements at small $x_B \sim 0.001$ and $Q^2\gtrsim 12$~GeV$^2$ have
been performed by the H1 collaboration \cite{Adloff} at HERA, and the
data presented in terms of the fragmentation invariant $\eta_h$.
The phenomenology of HMCs is markedly different in terms of $\eta_h$
from that discussed thus far in terms of $z_h$ because of the 
different functional forms for $\zeta_h$ in Eqs.~(\ref{eq:zeta_h}), 
which constrains $\zeta_h > \eta_h$, and because of the Jacobian
$d\zeta_h / d\eta_h$.
In their analysis of the H1 data, Albino {\it et al.} \cite{Albino}
included the $\zeta_h$ rescaling of the fragmentation process, but
neglected the effects of the target mass, which would be problematic
for heavier hadrons such as kaons and protons.
The H1 Collaboration measured charged hadron multiplicities,
dominated by pions ($\sim$60\%), with smaller contributions from
kaons ($\sim$30\%) and protons ($\sim$10\%).
In the measured $Q^2$ range the $m_h^2/Q^2$ term in $\xi_h$ is
therefore strongly suppressed and at the typically low $x_B$ values
one has $\xi \approx x_B$, so that overall we find the HMCs to be
similar to those in Ref.~\cite{Albino}.
However, for identified kaons, and especially protons, the SIDIS cross
section would be more strongly suppressed compared to the results of 
Ref.~\cite{Albino} because $\xi_h \approx x_B (1+m_h^2/Q^2)$ is 
significantly larger than $x_B$.
This suppression may be non-negligible for the extraction of kaon
and proton fragmentation functions from small-$x_B$ data.

%%%%%%%%%%%%%%%%%%%%%%%%%%%%%%%%%%%%%%%%%%%%%%%%%%%%%%%%%%%%%%%%%%%%%%%
\section{Conclusions}
\label{sec:conclusion}

In this paper we have derived hadron mass corrections to semi-inclusive
deep inelastic cross sections at finite $Q^2$ and have performed a 
systematic exploration of their phenomenological consequences.
Within the collinear factorization framework the modifications to the
leading order SIDIS cross sections from initial and final state masses
arise from a rescaling of the quark distribution and fragmentation
functions in terms of the modified Nachtmann scaling variable $\xi_h$
and a finite-$Q^2$ fragmentation variable $\zeta_h$, respectively.
The need for a modified Nachtmann variable is dictated by the
requirement that the physical kinematic thresholds for the
semi-inclusive process are explicitly respected.

We have examined the effects of the hadron mass corrections numerically
at kinematics relevant to recent experiments, finding sizable effects
at both small and large values of $z_h$, as well as for increasing 
$x_B$ and $m_h$, and low $Q^2$.
Our results emphasize the importance of controlling for such 
corrections in intermediate to high-$x_B$ experiments executed at
low $Q^2$, of which measurements at Jefferson Lab are typical,
although not exclusive.
We find that the hadron mass corrections can in some cases compete 
with the quoted experimental errors (as in the measurement of the
ratio of unfavored to favored fragmentation functions $D^-/D^+$) 
or overwhelm them (as in the cross section measurements). 
Due to the presence of the modified Nachtmann variable the HMCs may
also need to be considered for small-$x_B$ collider experiments for
the production of heavier hadrons such as kaons and protons, and 
their effects require further study.

The most direct use of the results presented here will be in
leading twist analyses of SIDIS cross sections, where the HMCs must
be included before extracting information on parton distribution
and fragmentation functions, especially at large $x_B$ and $z_h$.
Application of this work can also be found in studies of
semi-inclusive data in the nucleon resonance region, which has
been the focus of attention recently in view of understanding the
workings of quark-hadron duality \cite{MEK,ISGUR,CM}.

While the present analysis has been performed at leading order in
$\alpha_s$, in the future we plan to extend the formalism to
next-to-leading order, which will permit a more quantitative treatment of
transverse mass dependence of the produced hadrons, $p_{h\perp} \neq 0$.
It will also allow contact with transverse momentum dependent parton
distributions, in which nonzero parton transverse momentum,
$k_\perp \neq 0$, is an essential element.
Finally, as in the inclusive DIS case, the SIDIS cross section corrected
for hadron mass effects exhibits the threshold problem which renders it
nonzero as $x_B \to x_B^{\rm max}$.
Solutions of this problem proposed in the literature for inclusive
structure functions \cite{Tung,Steffens,AQ} will be extended to SIDIS
in future work.

%%%%%%%%%%%%%%%%%%%%%%%%%%%%%%%%%%%%%%%%%%%%%%%%%%%%%%%%%%%%%%%%%%%%%%%%%
\vspace*{0.5cm}
\begin{acknowledgments}

We thank S.~Albino, A.~Bacchetta, R.~Sassot and M.~Schlegel for helpful
discussions and communications.  This work was supported by the
DOE contract No. DE-AC05-06OR23177, under which Jefferson Science
Associates, LLC operates Jefferson Lab, and NSF award No.~0653508.
\end{acknowledgments}
\vspace*{0.5cm}

\appendix
%%%%%%%%%%%%%%%%%%%%%%%%%%%%%%%%%%%%%%%%%%%%%%%%%%%%%%%%%%%%%%%%%%%%%%%%%
\section{Collinear Frames in Semi-Inclusive DIS}
\label{sec:app}

A collinear frame in Minkowski space is defined by any two four-vectors.
The intersection of the plane where they lie with the light-cone defines
the light-cone four-vectors $\nbar^\mu$ and $n^\mu$, that satisfy
$n^2=\nbar^2=0$ and $\nbar \cdot n = 1$.
In SIDIS the hadronic tensor depends on the three vectors $p^\mu$,
$q^\mu$ and $p_h^\mu$, which define three possible collinear frames
denoted $(p,q)$, $(p_h,q)$ and $(p_h,p)$.
The $(p,q)$ frame is the only frame that can be defined in DIS
and is the one used in this work;
the $(p_h,q)$ frame is the only one that can be defined in
semi-inclusive hadron production in $e^+ e^-$ collisions;
and finally the $(p_h,p)$ frame is typically preferred for analysis
of transverse momentum dependent parton distributions in SIDIS.

In terms of the vectors $p$, $q$ and $p_h$ one can define two
fragmentation invariants,
\begin{align}
    z_h & = \frac{p_h \cdot p}{q \cdot p} \ , \qquad
  \eta_h = \frac{2 p_h \cdot q}{q^2}\, ,
\label{eq:zhzx}
\end{align}
which together with $x_B$, $Q^2$, $M^2$ and $m_h^2$ form a complete
set of scalar Lorentz invariants in SIDIS.
Because the variable $\eta_h$ is defined independently of the target
momentum, the effects of the final state hadron mass will decouple from
those of the target mass in all reference frames.
In contrast, $z_h$ is defined in terms of both the target and produced
hadron momenta, so that the target and hadron mass effects here will be
entangled.

The light-cone fractional momentum $\xi$ (Nachtmann variable) and
the fragmentation variable $\zeta_h$ are defined in terms of the
plus and minus components of the momenta as in Eqs.~(\ref{eq:xi})
and (\ref{eq:zeta_h1}),
\begin{align}
     \xi = - \frac{q^+}{p^+}\ ,   \qquad
 \zeta_h = \frac{p_h^-}{q^-} \ .
 \label{eq:xizetah}   
\end{align}
We use these definitions in all three frames; however, in each frame
the light-cone vectors (and therefore the plus and minus components of
the four-momenta) will be different.
In the following we discuss each of the three collinear frames and the
consequences within each frame of the choice of fragmentation invariant.

\vspace*{0.5cm}
% .....................................................................
{\bf $\bm{(p,q)}$ frame.}
In this frame the external vectors can be decomposed in terms of the
light-cone vectors $n$ and $\nbar$ as in Eqs.~(\ref{eq:kinematics}),
\begin{subequations}
\begin{eqnarray}   
  p^\mu   &=& p^+ \nbar^\mu
         + \frac{M^2}{2 p^+} n^\mu\ ,           \\
  q^\mu   &=& - \xi p^+ \nbar^\mu
         + \frac{Q^2}{2\xi p^+} n^\mu\ ,        \\
  p_h^\mu &=& \frac{m_{h\perp}^2}{\zeta_h Q^2/\xi} p^+ \nbar^\mu
         + \zeta_h \frac{Q^2}{2 \xi p^+} n^\mu
         + p_{h\perp}^{\,\mu} \ ,
\label{eq:kinematics_pq}
\end{eqnarray}
\end{subequations}
where $m_{h\perp}^2 = m_h^2 - p_{h\perp}^2 = m_h^2 + \bm{p}_{h\perp}^2$
is the transverse mass of the produced hadron.
Inverting the definition of $x_B$, the Nachtmann scaling variable
can be written as in Eq.~(\ref{eq:xi}),
\begin{align}
  \xi = \frac{2 x_B}{1 + \sqrt{1 + 4 x_B^2 M^2/Q^2}}\ .
\label{eq:xi_append}
\end{align}
Similarly, the hadron fractional momentum $\zeta_h$ can be expressed
in terms of either the fragmentation invariant $z_h$,
\begin{align}
  \zeta_h  = \frac{z_h}{2} \frac{\xi}{x_B}
    \left(
      1 + \sqrt{1 - 4 \frac{x_B^2}{z_h^2} \frac{M^2 m_{h\perp}^2}{Q^4}}
    \right)\ ,
  \label{eq:zeta_h-zh}
\end{align}
or in terms of $\eta_h$,
\begin{align}
  \zeta_h  = \frac{\eta_h}{2}
  \left( 1 + \sqrt{1 + 4 \frac{1}{\eta_h^2}
	 \frac{m_{h\perp}^2}{Q^2}}
  \right)\ .
  \label{eq:zeta_h-zx}
\end{align}
One can show that for any finite $\zeta_h$ the variable $z_h \to
\eta_h$ in the Bjorken limit. This is obviously true in any frame.

\vspace*{0.5cm}
% .....................................................................
{\bf $\bm{(p_h,q)}$ frame.}
In this frame, used in Ref.~\cite{Albino} for example, the external
SIDIS vectors are defined as
\begin{subequations}
\begin{eqnarray}
  p^\mu   &=& p^+ \nbar^\mu
         + \frac{M_\perp^2}{2 p^+} n^\mu
         + p_{\perp}^{\,\mu} \ ,                  \\
  q^\mu   &=& - \xi p^+ \nbar^\mu
         + \frac{Q^2}{2\xi p^+} n^\mu\ ,        \\
  p_h^\mu &=& \frac{m_{h}^2}{\zeta_h Q^2/\xi} p^+ \nbar^\mu
         + \zeta_h \frac{Q^2}{2 \xi p^+} n^\mu \ ,
\label{eq:kinematics_phq}
\end{eqnarray}
\end{subequations}
where $M_\perp^2 = M^2 - p_\perp^2 = M^2 + \bm{p}_\perp^2$ is the
transverse mass of the target nucleon.
The Nachtmann variable in this case is given by
\begin{align}
  \xi = \frac{2 x_B}{1 + \sqrt{1 + 4 x_B^2 M_\perp^2/Q^2}}\ ,
\end{align}   
which, in contrast to its definition in the $(p,q)$ frame,
depends explicitly on the transverse mass of the target nucleon.
Furthermore, in terms of the fragmentation invariant $z_h$,
the finite-$Q^2$ fragmentation variable $\zeta_h$ is given by
\begin{align}
  \zeta_h  = \frac{z_h}{2} \frac{\xi}{x_B}
    \left(
      1 + \sqrt{1 - 4 \frac{x_B^2}{z_h^2} \frac{M_\perp^2 m_h^2}{Q^4}}
    \right)\ ,  
  \label{eq:zeta_h-zh_phq} 
\end{align}
or in terms of $\eta_h$ by
\begin{align}
  \zeta_h  = \frac{\eta_h}{2}
    \left(
      1 + \sqrt{1 + 4 \frac{1}{\eta_h^2} \frac{m_h^2}{Q^2}}
    \right)\ .
  \label{eq:zeta_h-zx_phq}
\end{align}

\vspace*{0.5cm}
% .....................................................................
{\bf $\bm{(p_h,p)}$ frame.}
The external vectors in this frame, used in Ref.~\cite{Mulders} for
example, are given by
\begin{subequations}
\begin{eqnarray}
  p^\mu   &=& p^+ \nbar^\mu
         + \frac{M^2}{2 p^+} n^\mu \ ,         \\
  q^\mu   &=& - \xi p^+ \nbar^\mu
         + \frac{Q_\perp^2}{2\xi p^+} n^\mu
         + q_{\perp}^{\,\mu} \ ,                  \\
  p_h^\mu &=& \frac{m_{h}^2}{\zeta_h Q^2/\xi} p^+ \nbar^\mu  
         + \zeta_h \frac{Q^2}{2 \xi p^+} n^\mu \ ,
\label{eq:kinematics_php}
\end{eqnarray}
\end{subequations}
where $Q_\perp^2 = Q^2 - q_\perp^2 = Q^2 + \bm{q}_\perp^2$
is the transverse mass of the virtual photon.
The Nachtmann variable in this frame depends explicitly on $Q_\perp^2$,
\begin{align}
  \xi = \frac{Q_\perp^2}{Q^2}
  \frac{2 x_B}{1 + \sqrt{1 + 4 x_B^2 M^2 Q_\perp^2/Q^4}}\ ,
\end{align}
and the finite-$Q^2$ fragmentation variable is given by
\begin{align}
  \zeta_h  = \frac{z_h}{2} \frac{\xi}{x_B} \frac{Q^2}{Q_\perp^2}
    \left(
      1 + \sqrt{1 - 4 \frac{x_B^2}{z_h^2} \frac{M^2 m_h^2}{Q^4}}
    \right)\ ,
  \label{eq:zeta_h-zh_php}
\end{align}   
or
\begin{align}
  \zeta_h  = \frac{\eta_h}{2} \frac{Q^2}{Q_\perp^2}
    \left(
      1 + \sqrt{1 + \frac{4}{\eta_h^2} \frac{m_h^2 Q_\perp^2}{Q^4}}
    \right)\ .
  \label{eq:zeta_h-zx_php}
\end{align}

\vspace*{0.5cm}
% .....................................................................
{\bf Relations between frames.}
In general the frames discussed here are distinct.
However, to leading order in $1/Q^2$ the vectors $p$, $q$ and $p_h$
lie in the same plane and the three frames in fact coincide.
Comparing the $(p,q)$ and $(p_h,q)$ frames, for example, the
differences between the transverse momenta and scaling variables
can be expressed as
\begin{subequations}
\begin{eqnarray}
\bm{p}_{h\perp} &= \bm{p}_\perp^* + {\cal O}(\bm{p}_\perp^{*2}/Q^2)\, ,\\
\xi & = \xi^* + {\cal O}(\bm{p}_\perp^{*2}/Q^2)\, ,		       \\
\zeta_h & = \zeta_h^* + {\cal O}(\bm{p}_\perp^{*2}/Q^2)\, ,
\end{eqnarray}
\end{subequations}
where the asterisks ($^*$) label quantities in the $(p_h,q)$ frame.
Similar relations are applicable also for the parton fractional 
momentum $x$ and the hadron fractional momentum $z$.
At leading order in collinear factorization one has
$\bm{p}_{h\perp} = 0$ and the frames are manifestly equivalent.
Moreover, since $\langle \bm{p}^2_{h\perp} \rangle \ll Q^2$ for
$\bm{p}_{h\perp}$-integrated cross sections at next-to-leading order
the differences between the collinear frames should remain small.  
It will nevertheless be important to check whether, and in what 
kinematic range, this approximation is valid.
On the other hand, for unintegrated cross sections the differences
between frames become relevant and their effects must be quantified.

%%%%%%%%%%%%%%%%%%%%%%%%%%%%%%%%%%%%%%%%%%%%%%%%%%%%%%%%%%%%%%%%%%%%%%%

\end{document}